\documentclass{article}

\usepackage{arxiv}

\usepackage[utf8]{inputenc} 
\usepackage[T1]{fontenc}    
\usepackage{hyperref}       
\usepackage{url}            
\usepackage{booktabs}       
\usepackage{amsmath}
\usepackage{float}

\usepackage{amsfonts}       
\usepackage{nicefrac}       
\usepackage{microtype}      
\usepackage{lipsum}		
\usepackage{graphicx}
\usepackage[authoryear]{natbib}			
\usepackage[nameinlink, capitalize]{cleveref} 
\usepackage{etoolbox} 
\usepackage{epstopdf} 

\usepackage{caption}
\usepackage[flushleft]{threeparttable}
\usepackage{tabularx}
\usepackage{multirow}

\makeatletter

\patchcmd{\NAT@test}{\else \NAT@nm}{\else \NAT@nmfmt{\NAT@nm}}{}{}

\DeclareRobustCommand\citepos
  {\begingroup
   \let\NAT@nmfmt\NAT@posfmt
   \NAT@swafalse\let\NAT@ctype\z@\NAT@partrue
   \@ifstar{\NAT@fulltrue\NAT@citetp}{\NAT@fullfalse\NAT@citetp}}

\let\NAT@orig@nmfmt\NAT@nmfmt
\def\NAT@posfmt#1{\NAT@orig@nmfmt{#1's}}

\makeatother

\title{Inflation Dynamics of Financial Shocks}

\author{ \href{https://orcid.org/0000-0000-0000-0000}{Olli Palmén}\\
	Faculty of Social Sciences\\
	University of Helsinki\\
	P.O. Box 17 (Arkadiankatu 7), \\
	00014 University of Helsinki, Finland \\
	\texttt{olli.palmen@helsinki.fi} \\
}

\renewcommand{\headeright}{Working Paper}
\renewcommand{\undertitle}{Working Paper}

\hypersetup{
pdftitle={Inflation dynamics of financial shocks},
pdfsubject={},
pdfauthor={Olli Palmén},
pdfkeywords={Inflation, financial shocks, structural vector autoregression},
}

\begin{document}
\maketitle

\begin{abstract}
We study the effects of financial shocks on the United States economy by using a Bayesian structural vector autoregressive (SVAR) model that exploits the non-normalities in the data. We use this method to uniquely identify the model and employ inequality constraints to single out financial shocks. The results point to the existence of two distinct financial shocks that have opposing effects on inflation, which supports the idea that financial shocks are transmitted to the real economy through both demand and supply side channels.
\end{abstract}


\section{Introduction} \label{sec:introduction}
Financial shocks have garnered considerable attention in the macroeconomic literature following the Global Financial Crisis. Drawing from the recent experience, a number of studies find that disturbances in the financial sector play a significant role in explaining business cycle fluctuations. However, notwithstanding the considerable attention to the role of the financial sector, the inflationary effects of financial shocks are not obvious. 

According to macroeconomic theory financial shocks may have both inflationary and disinflationary effects. Expansionary financial shocks are generally found to lead both to an increase in output and inflation through an increase in households' and firms' demand \citep{Curdia2010, Gertler2011}. However, \cite{Gerali2010} observe that the inflationary response of financial shocks may depend on the specific type of the shock. That is, the sign of the impact effect on inflation is different for shocks that affect banks' capital and those that have an effect on the supply of credit, such as household borrowing conditions. More recently, \cite{Gilchrist2017} argue that financial market frictions attenuate the effect of output on prices, which explains the relatively mild deflationary pressures experienced during the Great Recession of 2008.

Empirical estimates on the inflation dynamics of financial shocks are also not clear. Recent studies that use sign-identified structural vector autoregressive (SVAR) models to estimate the effects of financial shocks generally assume a positive co-movement between inflation and output as predicted by theory \citep{Gambetti2017}. However, to account for the fact that theoretical macroeconomic models are consistent with both negative and positive co-movement between output and inflation, recent empirical studies choose to remain agnostic about the effect of financial shocks on inflation (see for example, \cite{Abbate2016} and \cite{Meinen2018}). Nevertheless, these studies provide evidence in favor of both negative and positive co-movement between output and inflation.

In this paper, we re-examine the inflation effects of financial shocks using a Bayesian SVAR model by \cite{Lanne2020} that exploits the non-normalities in the data. This method allows us to statistically identify the model only based on the statistical properties of the time series. Given that statistically identified shocks do not have an economic interpretation, we use theoretical predictions about the dynamics of the impulse responses and other extraneous information to label the shocks. Our method stands in contrast to the traditional method of using sign restrictions to identify SVAR models in two important ways: First, our approach achieves point-identification, whereas the sign identification procedure provides a set of admissible models, which are consistent with the restriction and data \citep{Baumeister2015}. Second, our identification approach does not require any restrictions on the model parameters, whereas sign-identification requires restrictions on the impulse responses. Moreover, our approach does not guarantee that a shock (or shocks) satisfying a set of inequality constraints on the impulse responses exists. Our approach also allows us to assess the whether the data lend support to sign restrictions, which have been used in previous empirical studies.

Our results point to the existence of two distinct financial shocks that have opposing effects on inflation. This finding supports the idea that financial shocks are transmitted to the real economy through both demand and supply side channels. This suggests that theoretical and empirical studies on financial shocks would benefit from the inclusion of both demand and supply channels of financial shocks to avoid problems with misspecification. The results are also relevant from a policy perspective, given that cost-push shocks, which increase inflation irrespective of aggregate demand, pose a trade-off between central banks' dual objective of price and output stability.

The rest of the paper is organized as follows. \cref{sec:literature} presents an overview on the related literature. \cref{sec:methods} explains the methodology and the data.  \cref{sec:results} presents and discusses results. \cref{sec:conclusion} concludes.

\section{Literature}  \label{sec:literature}

Financial shocks in theoretical models are compatible with both inflationary and disinflationary effects, depending on the transmission channel and the central banks' policy rule. In general, the macroeconomic literature finds that financial shocks affect prices through the demand channel. \cite{Gertler2011} show that an unexpected decrease in the bank capital leads to a fall in inflation by reducing firms' and households' demand for both investment and consumption. However, the central banks' response to tensions in the credit markets may offset the disinflationary effect of financial shocks. In \citepos{Curdia2010} model, a standard Taylor rule that weights output and inflation leads to a positive co-movement between prices and output, whereas a policy rule that takes into account the tightening of private sector borrowing conditions has the opposite effect.

More recently, several studies emphasize the role of credit markets in explaining the unexpectedly modest fall in inflation during the Great Recession using DSGE models. In \citepos{Gilchrist2017} model, financially constrained firms set prices above their marginal cost during financial crises in order to avoid accessing costly external finance, which leads to opposite movements in inflation and output. \cite{Meh2010} find that banks' financial position plays an important role in explaining macroeconomic fluctuations, given that bank capital mitigates the moral hazard problem between bankers that finance projects and investors who supply the funding. In their model, an unexpected negative shock to bank capital leads to a protracted decrease in lending and investment. However, aggregate consumption and inflation slightly increase following the shock, because the subsequent increase in the price of capital goods cause households to increase the consumption of final goods. Financial shocks also behave in the same way as cost-push shocks in the model of \cite{DeFiore2013}, where an increase in borrowing costs drive up firms' marginal costs, which are subsequently passed to consumers.

Theoretical models are also compatible with financial shocks whose inflationary effects depend on the origin of the shock. In \citepos{Gerali2010} model, a shock that decreases bank capital induces banks to raise the price of lending, which, in turn, reduces firms' investment demand. However, an increase in firms' demand for labor drives up wages, and therefore induces a negative co-movement between output and prices. On the other hand, another type of shock originating in the financial sector, characterized by an unexpected fall in banks' loan-to-value ratios, has a negative effect on output and inflation\footnote{The response of inflation to loan-to-value shocks is based on additional information provided in \cite{Gambetti2017}}.

A number of recent empirical studies use SVAR models to identify and estimate the macroeconomic effects of financial shocks and assess their contribution to business cycle fluctuations during the recent economic and financial crises. Sign restrictions have been commonly used to identify financial shocks in these studies. The restrictions are generally based on the predicted effect of shocks in New Keynesian dynamic stochastic general equilibrium (DSGE) models that include a financial sector. However, inconsistencies in the responses of key variables to financial shocks in DSGE models pose an inherent difficulty in assigning sign restrictions. To overcome this problem, sign restrictions have been commonly based on features, with which a set of benchmark DSGE models are generally in agreement (see, for example, \citep{Gambetti2017}).

Empirical studies on financial shocks commonly impose a positive co-movement between output and inflation. This restriction is largely motivated mainly by theoretical considerations \citep{Busch2010, Darracq-Paries2015} but also provides a method to distinguish financial shocks from aggregate supply shocks, which are characterized as having an opposite effect on output and prices \citep{Furlanetto2017, Gambetti2017}. The restrictions are commonly imposed on the period during which the shock occurs, but sometimes only on further lags to account for the delayed response of real macroeconomic variables. For instance, \cite{Busch2010} impose no restrictions on the impact response of output and prices but set a positive co-movement for these variables for two periods after the initial response. Although the sign of the response of is not restricted before two periods, \citeauthor{Busch2010} find that financial shocks also exhibit positive co-movement between output and prices upon impact.

In other empirical studies, assumption about the timing of the impact of shocks are used to distinguish real economic shocks, such as shocks affecting aggregate demand or aggregate supply, from financial shocks. \citep{Barnett2014} imposes restrictions on the financial variables on impact following a credit supply shock, but sets the responses of macroeconomic variables such as prices to be zero on impact and unrestricted for further lags  to account for the time that it takes for financial shocks to be transmitted to the real economy. Moreover, real economic shocks are assumed to have an immediate impact on the output and prices, whereas the impact on financial variables is unrestricted. \cite{Peersman2011b} also identifies bank lending shocks by imposing a zero restriction on the real variables on impact and a lagged effect on bank lending, whereas restriction on the financial  variables are imposed on impact. \cite{Peersman2011b} reports a positive co-movement between output and inflation at longer horizons in the Euro area, whereas \cite{Barnett2014} find that expansionary financial shocks have a negative impact on inflation in the United Kingdom. \cite{Gilchrist2012} impose a recursive structure on the shocks such that unexpected shocks in the financial markets have no immediate impact on real economic variables. For the United States, they find a non-negligible impact on inflation three quarters after the shock and a positive co-movement between output and inflation at longer horizons.

Several recent studies attempt to reconcile the opposing predictions in the theoretical models by remaining agnostic about the sign of inflation following financial shocks. \cite{Abbate2016} disentangle financial shocks from other structural shocks by sign and exclusion restrictions, while imposing no restriction on the sign of the inflation. They find the contractionary financial shock to have inflationary effects, which supports the hypothesis that financial shocks partly explain the missing disinflation following the financial crisis. These findings are somewhat at odds with those of \cite{Hristov2012} who observe a negative, albeit non-significant, effect on inflation after expansionary financial shocks in a panel of euro area countries. Moreover, \cite{Meinen2018} report an ambiguous effect on inflation following financial shocks  in both the euro area and the United States.

This paper is related to \cite{Brunnermeier2017} who study the macroeconomic effects of financial stress in a structural VAR model. They similarly use statistical properties of the model to identify the structural shocks, allowing for heteroskedasticity and non-normality of the residuals. Their findings complement our results, in that they succeed in identifying two distinct shocks that capture tensions in the financial markets that in their effects on prices and credit aggregates. In contrast to \citeauthor{Brunnermeier2017}, we document negative co-movement between prices and output for one type of credit-side shock.

\section{Model and estimation}  \label{sec:methods}

The objective of this paper is to study the inflationary impact of financial shocks in the US. To that end, we estimate a Bayesian VAR model with non-normal errors by \cite{Lanne2020}, who apply the method to the crude oil market and monetary policy shocks. The method has also been previously used  to study the effects of unconventional monetary policy \citep{Puonti2019}. The model exploits non-normalities in the data to uniquely identify the model, as proposed by \cite{Lanne2017}. The advantage of this approach is that restrictions on the model parameters are not needed for the identification or model inference. However, extraneous information, for example from theoretical models, are needed to give an economic interpretation of the shocks. The labelling of the shocks is discussed in detail in \Cref{sec:identification}.

\subsection{The SVAR model with non-normal errors}

The structural VAR model is of the form

\begin{equation}
y_{t} = a_0 + \sum_{l=1}^{p }A_{l=1} y_{t-l} +  B \epsilon_{t}.
\label{svar}
\end{equation}

$y_{t}$ is an $N \times 1$ vector of endogenous variables. $a_0$ is an $N \times 1$ vector of constants and $A_{l}$ ($l=1,\ldots, p$) are the $N \times N$ coefficient matrices for the $l$th lag of the endogenous variables. The $N \times N$ matrix $B$ captures the contemporaneous relations between the structural shocks. The error process $\epsilon_{t} = [\epsilon_{1,t}, \ldots,  \epsilon_ {n,t}]$ is a $N \times 1$ vector of independently distributed random vectors. Moreover, in contrast to the common assumption of Gaussian errors, we assume that the $i$th component $\epsilon_{i,t}$, $i=1,\ldots,N$  follows a $t$-distribution with degrees of freedom $\lambda_i>2$. 

We assume that the SVAR process is stable such that

\begin{equation}
\text{det} \left( I_n - A_1 z - \cdots - A_p z^p \right) \neq 0, 	\quad \vert z \vert  \leq 1.
\end{equation}

and therefore the process has a moving average representation

\begin{equation}
y_t = \mu + \sum_{j=0}^{\infty} \Psi_j B \epsilon_{t-j} = \mu + \sum_{j=0}^{\infty} \Theta_j \epsilon_{t-j},
\end{equation}

where $\mu = A(1)^{-1}a_0$ and $A(L)=I_K - A_1L - \cdots - A_p L^p$ and the $N \times N$ coefficient matrices $\Theta_j$ can be obtained recursively

\begin{align*}
\Theta_0 &= B \\
\Theta_j &= \sum_{l=1}^j \Theta_{j-l}A_l, \quad j=1,2,\ldots,
\end{align*}

where $A_l = 0$ for $l>p$. The $i$th column of $\Theta_j$, contains the impulse responses of the endogenous variables to the $i$th structural shock $\epsilon_{i,t}$, $i=1,\ldots,N$.

As shown by \cite{Lanne2017}, given the assumptions about the non-normality and independence of error term $\epsilon_t$ hold matrix and that at most one of the structural shocks is normally distributed, the SVAR model (\ref{svar}) is point-identified apart from permutation and scaling of the columns of matrix $B$\footnote{For a detailed description of the identification see Proposition 1 in \cite{Lanne2017}}. Unique identification is achieved by choosing a single model from $n!$ observationally equivalent set of models using the algorithm described in \cite{Lanne2017}.

\subsection{Model priors and specification}

The model estimation and the choice of priors closely follow \cite{Lanne2020}, who provide a detailed description of the model and the estimation method. Next, we provide an overview of the model priors and the specification of the model.

We operate on the inverse of matrix $B$, and define a prior for $B^{-1}$ such that 

\begin{equation}
b \sim \mathcal{N}(\underline{b}, \underline{V}_b),
\end{equation}

where $b \equiv \text{vec}(B^{-1})$.

Moreover, we the prior distribution for each $\lambda_i$ is assumed to be exponential such that

\begin{equation}
\lambda_i \sim \text{exp}(\underline{\lambda}_i).
\end{equation}

The prior for $a \equiv	\text{vec}(A')$ is assumed to be multivariate normal:

\begin{equation}
a \sim \mathcal{N}(\underline{a}, \underline{V}_a).
\label{aprior}
\end{equation}

We assume a Minnesota-type prior for the model parameters such that the prior variance for the \textit{pq}th element of $A_l$ ($l=1,\ldots, p$) is $v_{pq,l} = (\kappa_1/l^\kappa_3)^2$ if $p=q$ and $v_{pq,l} = (\kappa_1 \kappa_2 \sigma_p/l^\kappa_3 \sigma_q)^2$ if $p\neq q$. The prior variance for the constant term $a_0$ is assumed to be $(\sigma_p \kappa_4)^2$.

The Minnesota prior for the lagged coefficients of the VAR in \cref{aprior} is set up such that $\kappa_1 = 3$, $\kappa_2 = 0.5$, $\kappa_3 = 1$. The prior variance for the constant is assumed to be non-informative such that $\kappa_4 = 100$. Given that the macroeconomic data enter the model in first differences, we set the first the coefficient for the first lag of each endogenous variable equal to zero. We set the prior mean $\underline{\lambda}_i=10$ for each degree of freedom parameter $\lambda_i$, such that the prior is close to a normal distribution. The prior for each element in $b$ has mean zero and standard deviation $10^3$, which results in a non-informative prior for  $B^{-1}$.

The lag length of the VAR is specified to be two, based on the Akaike information criterion with maximum lag length of 10. The results are based on a sample of 1000000 draws after a burn-in of 100000 iterations using the Gibbs sampler described in the appendix of \cite{Lanne2020}.

\subsection{Data}

The dataset includes quarterly data for the United States, covering the period from 1980Q1 to 2011Q4\footnote{The dataset is obtained from \cite{Gambetti2017}, who also provides a detailed description of the data. The same dataset is used for the comparability of the results. The dataset was downloaded from the Journal of Applied Econometrics Data Archive (\url{http://qed.econ.queensu.ca/jae/2017-v32.4/})}. The endogenous variables included in the model cover the real and monetary sectors of the economy to capture the role of financial shocks on business cycles: real GDP, consumer price index, lending to households on non-financial corporations, the composite lending rate, the three-month Treasury Bill rate.

\subsection{Labelling of financial shocks}
\label{sec:identification}

In this section we discuss the economic labelling of the financial shocks. Because the model identification is purely based on the statistical properties of the data, extraneous information is needed to give the shocks an economic interpretation. As discussed above, in theoretical literature, a financial shock is both compatible with a positive \citep[e.g.,][]{Gertler2011} and a negative \citep[e.g.,][]{Gilchrist2017} co-movement with output and inflation or more than one financial shock that vary in their inflationary response \citep{Gerali2010}. Based on this observation, in contrast to previous empirical research, we consider the possibility that there exist two financial shocks with opposite inflationary responses. 

We proceed to label to shocks  by assessing the posterior probability that the structural shocks satisfy a set of inequality constraints based on theoretical predictions\footnote{Although we assign probabilities to shocks based on the sign of the impulse responses, our leads to point identification, whereas the traditional sign-restriction approach provides an admissible set of models.}. To this end, we adopt a set of inequality constraints that have been used in previous studies that use sign restrictions to identify financial shocks \citep[e.g.,][]{Gambetti2017, Hristov2012}. The inequality constraints are summarized in \Cref{tab:restrictions}. The shocks are normalized to have a positive impact on the lending rate\footnote{The identification is based on the sign pattern of the shocks, and therefore the normalization of the sign of the shock has no real implication on the identification procedure.}. 

We are interested in two distinct financial shocks with opposing inflationary responses. Because the inflationary response generally depends on the transmission channel of the shock, we label these shocks as supply side and demand side financial shocks, supply side shocks being those that exhibit negative co-movement between output and inflation, whereas the correlation is positive for demand side shocks.

\begin{table}[H]
\small
  \begin{threeparttable}
  \centering
    \caption{Inequality restrictions for identification}
      \label{tab:restrictions}
      \renewcommand{\arraystretch}{0.9}
     \begin{tabularx}{\textwidth}{lccccc}
        \toprule
              & \multicolumn{5}{c}{\textit{Variable}}\\
          		& \multicolumn{1}{p{1.6cm}}{\centering Real GDP growth}&  \multicolumn{1}{p{1.6cm}}{\centering Inflation}&  \multicolumn{1}{p{2cm}}{\centering Loan volume growth} &  \multicolumn{1}{p{1.6cm}}{\centering Lending rate}&  \multicolumn{1}{p{2cm}}{\centering Short-term interest rate} \\
        \midrule
     \multicolumn{1}{p{5.2cm}}{Demand side financial shock ($R_1$)}  & $-$ & $-$ & $-$ & $+$ & $-$ \\
          \multicolumn{1}{p{5.2cm}}{Supply side financial shock ($R_2$)}  & $-$ & $+$ & $-$ & $+$ & $-$ \\
        \bottomrule
     \end{tabularx}
       \captionsetup{justification=centering}
    \begin{tablenotes}
      \small
      \item \textit{Notes:} $+$/$-$ denote the sign of impact of impulse responses for each variable. The restrictions are imposed on the impulse responses on impact for all variables.
    \end{tablenotes}
  \end{threeparttable}

\end{table}

As is common in the literature, we assume that recessionary financial shocks increase the composite lending rate and decrease both the real GDP growth and the growth of bank lending within the same quarter. However, these properties are not sufficient to disentangle financial shocks from aggregate supply shocks, which are generally characterized by negative co-movement between output and prices. \cite{Gambetti2017} separate financial and aggregate supply shocks by imposing a positive co-movement between inflation and real GDP growth following financial shocks, while only imposing restrictions on the response of inflation and GDP growth to aggregate supply shocks. By contrast, \cite{Hristov2012} remain agnostic about the sign of consumer price inflation in response to a financial shock. However, in order to distinguish aggregate supply shock from financial shocks, \citeauthor{Hristov2012} assign a positive sign for the short-term interest rate in response to a contractionary aggregate supply shock. This implies that in the absence of financial tensions the monetary authority increases interest rates in response to higher inflation, despite a fall in output.

We assume that the response of the short-term interest rate to financial shocks is negative, which is underpinned by the assumption that the monetary policy authority responds immediately to negative financial shocks independent of the inflationary response. This is consistent with the optimal monetary policy response to financial shocks as described in \cite{DeFiore2013}. This also implies that, in addition to its price stability objective, the monetary authority responds to contractionary financial shocks by lowering its policy rate despite inflationary pressures. 

The inference on the structural shocks is based on the posterior distribution of the parameters of interest. The shocks are given an economic label by assessing whether the impulse response functions satisfy a set of inequality constraints based on theoretical predictions. In practice, this is done and calculating the posterior probabilities of each pair of shocks satisfying these constraints\footnote{Although we identify two shocks, the posterior probabilities can be calculated for any number of shocks $g$, for $1 \leq g \leq N$, as described in \cite{Lanne2020}}.

The inequality constraints with respect to two structural shocks are collected in the $N \times M$ matrix $R_m$ ($m=1,2$), where $N=5$ is the number of variables and $M=5$ is the number of constraints imposed on each shock $m$. Each row of $R_m$ represents one inequality constraint and contains only zeros, and a single entry with $1$ or $-1$.

Given the constraints described in \Cref{tab:restrictions}, we collect the inequality constraints in the matrices $R_m$, such that

\begin{equation}
R_1 = 
  \begin{bmatrix}
    -1 & 0 & 0 & 0 & 0 \\
    0 & -1 & 0 & 0 & 0 \\
    0 & 0 & -1 & 0 & 0 \\
    0 & 0 & 0 & 1 & 0 \\
    0 & 0 & 0 & 0 & -1
  \end{bmatrix} \quad \text{and} \quad
R_2 = 
  \begin{bmatrix}
    -1 & 0 & 0 & 0 & 0 \\
    0 & 1 & 0 & 0 & 0 \\
    0 & 0 & -1 & 0 & 0 \\
    0 & 0 & 0 & 1 & 0 \\
    0 & 0 & 0 & 0 & -1    \end{bmatrix}. 
\end{equation}

Let $B_i$ be the $i$th column of the matrix $B$, that is the column of the structural impact matrix corresponding to structural shock $\epsilon_{i,t}$ ($i=1,\ldots, N$)\footnote{The inequality constraints can be extended to a longer horizon of the impulse responses by applying constraints on the elements of $\Theta_j$, $j>0$. See \cite{Lanne2020}.}. 

The columns of the structural impact multiplier matrix $B$ that satisfy the inequality constraints are then included in the set $Q_m$, such that

\begin{equation}
Q_m = \left\lbrace B_{i} : R_m B_{i} \geq 0_{M \times 1} \right\rbrace.
\end{equation}

Then, for each pair of structural shocks $\epsilon_{i,t}$ and $\epsilon_{k,t}$, for $i,k = 1,\ldots,N,$ and $i \neq k$, the posterior probability of the constrained model conditional on the data is 

\begin{equation}
\text{Pr} \left( B_{i} \in Q_1, B_{k} \in Q_2, B_{m \neq i,k} \in Q^C \vert y \right), 
\label{eq:prob} 
\end{equation}

where $Q^C$ denotes the complement of the union $Q_1 \cup Q_2$. 

In practice, we first normalize one row of $B$ in all posterior draws such that the impact effects on the endogenous variables are compatible with one of the constraints collected in matrix $R_j$. In this case, we normalize the response of real GDP growth to be negative on impact for each structural shock. Then the posterior probabilities are calculated by counting the share of posterior draws for which each of the $N(N-1)$ pairs of shocks satisfy the constraints and no other shock satisfies either of the constraints.


Based on the posterior probabilities, we then calculate the Bayes factors for each pair of shocks against the unconstrained model to assess the plausibility of the data supporting the inequality constraints. The Bayes factor is given as the ratio of the marginal likelihoods of the constrained model ($M_r$) against the unconstrained model ($M_u$) is given by

\begin{equation}
\frac{p(y\vert M_r)}{p(y\vert M_u)}  = \frac{\int_S p(y\vert \phi) p(\phi) d\phi}{\int_{\Phi} p(y\vert \phi) p(\phi) d\phi} \frac{1}{\int_S p(\phi) d\phi},
\end{equation}

where $\phi$ is a vector of model parameters $p(\phi)$ is the prior distribution, and $p(y\vert \phi)$ is the marginal likelihood. $S$ denotes the set of values of $\phi$ for which the inequality constraints are satisfied.

Given that the identification is based on the statistical properties of the time series, it is possible that none, only one or more than one pair of shocks is consistent with these constraints. If there is no support for any of the shock pairs to satisfy the inequality constraints, we can conclude that there are does not exist two financial shocks with distinctive inflationary responses. In this case, we assess if a single financial shock with either positive or negative correlation with output and inflation exists\footnote{Model inference for a single shock is discussed in detail in \cite{Lanne2020}.}. If the data supports only one pair of shocks over the others, we may infer that there are two distinct financial shocks with opposite effects on inflation. If more than one pair of shocks are supported by the data, additional information is needed to distinguish between the shocks of interest.

Following \cite{Kass1995}, we consider a Bayes factor exceeding 3.2 supportive of the constrained model in favor of the unconstrained model. If only one constrained model exceeds the threshold, the pair of structural shocks are labelled accordingly. If the Bayes factor for none of the constrained models exceeds 3.2, the data is not considered supportive of the inequality constraints. However, it is also possible that more than one constrained model has Bayes factor exceeding the threshold. In this case we calculate the Bayes factor comparing the two constrained models. If, in this case, the Bayes factors exceeds 3.2, we consider that model the one most likely to satisfy the inequality constraints. If the Bayes factors do not exceed the threshold, other criteria are necessary to label the structural shocks.

\section{Results}  \label{sec:results}

In this section, we discuss the identification of the shocks and their macroeconomic implications on the US economy.

\subsection{Shock identification}

\begin{figure}[H]
\caption{Posterior densities of $\lambda_i$}
\includegraphics[scale=1]{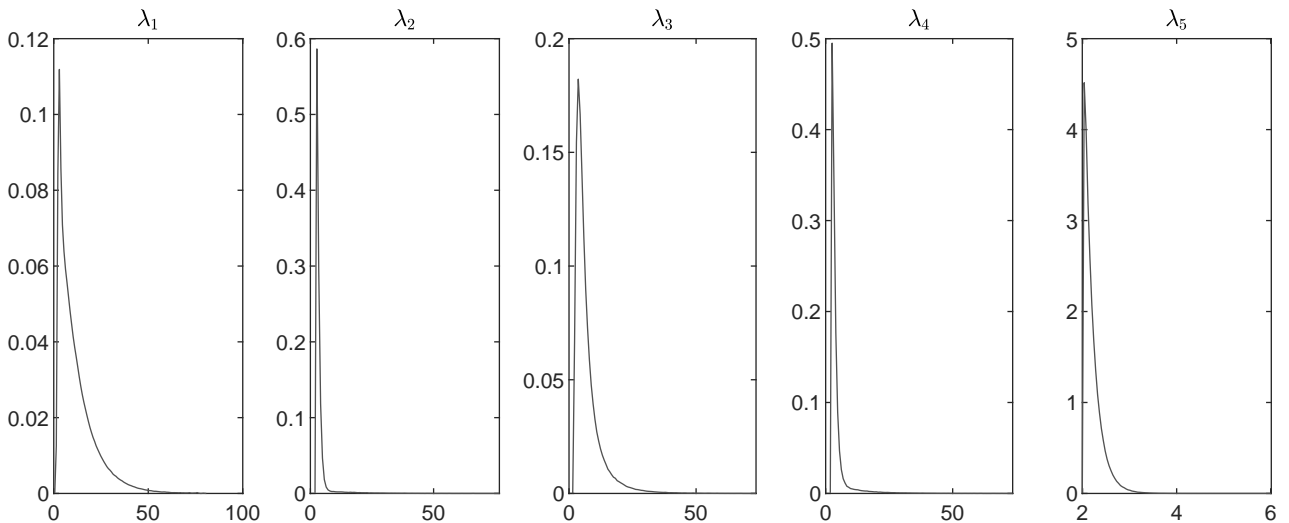}
\label{fig:lambda}
\end{figure}

Given that the identification of the shocks depends on the assumption that the structural errors follow a $t$-distribution and at most one of them is Gaussian, we first examine the posterior densities of the degree of freedom parameters $\lambda_i$. As can be seen in \Cref{fig:lambda}, the posterior distribution of $\lambda_i$ is centered around small values, with the mean ranging between 2.22 and 11.7. This implies that the distribution of the structural shocks is indeed heavy-tailed, thus satisfying the distributional assumptions for identification.

With the objective of labelling the shocks, we calculate posterior probabilities and corresponding Bayes factors to assess the plausibility that the structural shocks satisfy both of the inequality constraints, as detailed in \Cref{tab:restrictions}. The Bayes factors for the $N(N-1)$ pairs of the shocks against the unconstrained SVAR model are reported in \Cref{tab:BF}.


\begin{table}[H]
\begin{threeparttable}
\caption{Bayes factors for the combination of shocks satisfying the inequality restrictions}
\smallskip
\label{tab:BF}
\renewcommand{\arraystretch}{1.2}
\renewcommand\tabcolsep{18pt}

\begin{tabular}{cc|ccccc}
\toprule
\multicolumn{2}{c}{} & \multicolumn{5}{c}{$R_2$} \\ 
& &$\epsilon_1$  &$\epsilon_2$ &$\epsilon_3$ &$\epsilon_4$& $\epsilon_5$ \\
\cline{2-7}
\multirow{5}*{$R_1$} 
& $\epsilon_1$ & - & 0.01 & 0 & 0 & 0 \\
& $\epsilon_2$ & 1.55 & - & 13.24 & 0.01 & 2.58 \\
& $\epsilon_3$ & 0.19 & 0.01 & - & 0 & 0.155 \\
& $\epsilon_4$ & 0 & 0 & 0 &- & 0 \\
& $\epsilon_5$ & 0.06 & 0.20 & 3.60 & 0 & - \\
\bottomrule
\end{tabular}
    \begin{tablenotes}
      \small
      \item \textit{Notes:} The values denote Bayes factors for the $N(N-1)$ combinations of the columns of $B$, given the inequality constraints $R_1$ and $R_2$. 
    \end{tablenotes}
\end{threeparttable}
\end{table}

The data provides evidence for two of the shock pairs ((2,3) and (5,3)) satisfying the restrictions with  Bayes factors of 13.3 and 3.6, respectively. Moreover, there is weak or no support for other shock pairs since the Bayes factors for all other combinations are below the threshold of 3.2\footnote{The results are robust to applying the restrictions to impulse responses four quarters following the shock.}. For both of these models, $\epsilon_3$ satisfies the inequality constraints for the demand side financial shock and both $\epsilon_2$ and $\epsilon_5$ satisfy the constraints for the demand side financial shock. However, comparison of the two constrained models lends strong support to the shock pair (2,3) over shock pair (5,3) on the basis that the Bayes factor comparing the two constrained model is 3.6. Therefore, we label shocks $\epsilon_3$ and $\epsilon_2$ as the supply side and demand side financial shock, respectively.

The data therefore lends support to the existence of two financial shocks that vary in their inflationary response. This suggest that financial shocks are transmitted to the real economy via both aggregate demand and aggregate supply channels, which is consistent with predictions in the theoretical literature \citep[see e.g.,][]{Gerali2010}. \cite{Brunnermeier2017} also find two distinct shocks relating to financial stress, one relating to unexpected increases in bond spreads and the other relating to inter-bank lending. However the bond spread shock and the inter-bank shock both lead to positive co-movement between output and prices, although bond spread shock has a more persistent effect on output and business credit.

The existence of these shocks also has implications for the use of sign restrictions to identify financial shocks in structural VAR models. The conventional method of restricting the sign of the response of inflation to financial shocks, as employed by, for example, \cite{Gambetti2017} and \cite{Barnett2014} will only capture a particular transmission channel of financial shocks, and might therefore, downplay their macroeconomic importance. Moreover, the decision to remain agnostic about the sign of inflation needs to supplemented with extraneous information that may be used to disentangle supply and demand side channels of financial shocks. Finally, given that  a supply side financial shock and an aggregate supply shock both have opposing effects on prices and output, imposing restrictions only on these two variables is not sufficient to distinguish between these two shocks. Therefore, for example assumptions about the timing of the effects of these two shocks might be considered to disentangle the shocks.

\subsection{Macroeconomic implications of financial shocks}

\begin{figure}[H]
\caption{Impulse response functions to a demand side financial shock}
\smallskip
\includegraphics[scale=1]{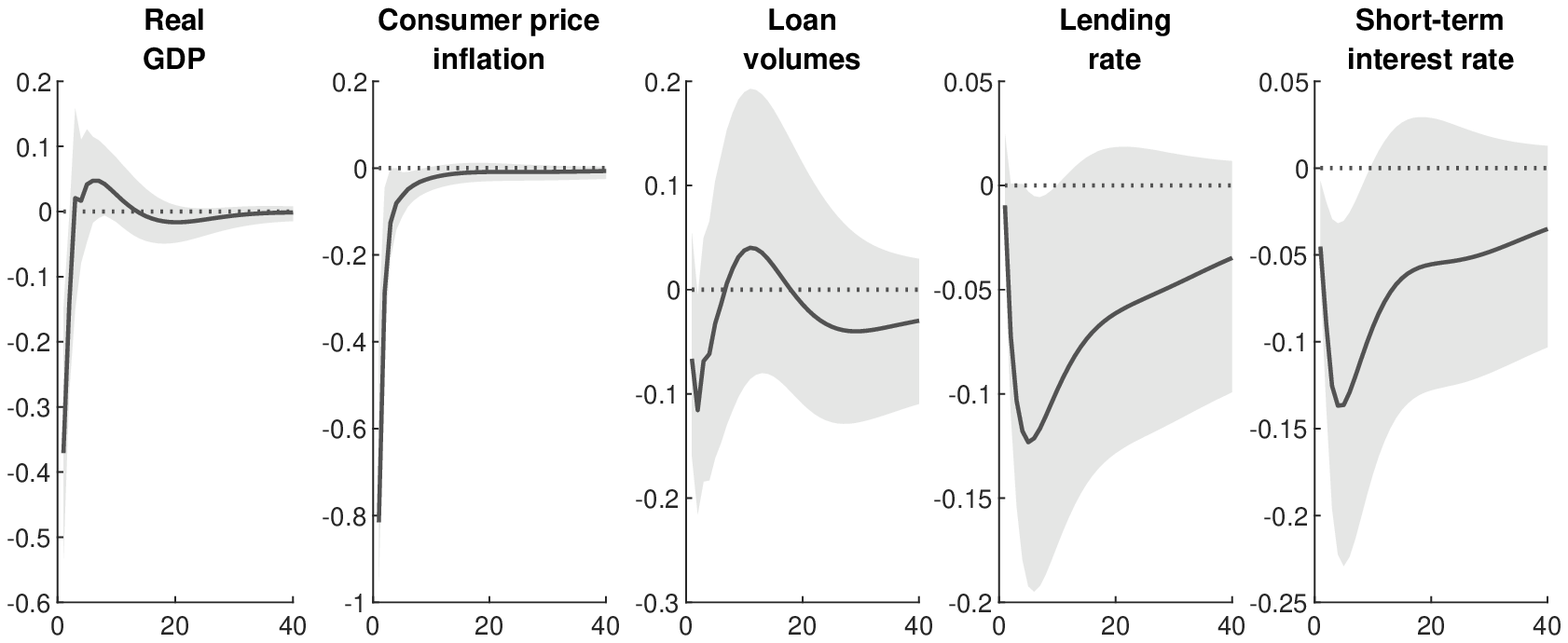}
\smallskip
\caption*{\textit{Notes}: The solid line is the posterior median and dashed lines are the 16th and 84th percentile of the posterior probability. Model is a VAR(2) including a constant. The estimation period is 1980:Q1-2011:Q4}
\label{fig:irf2}
\end{figure}

\begin{figure}[H]
\caption{Impulse response functions to a supply side financial shock}
\smallskip
\includegraphics[scale=1]{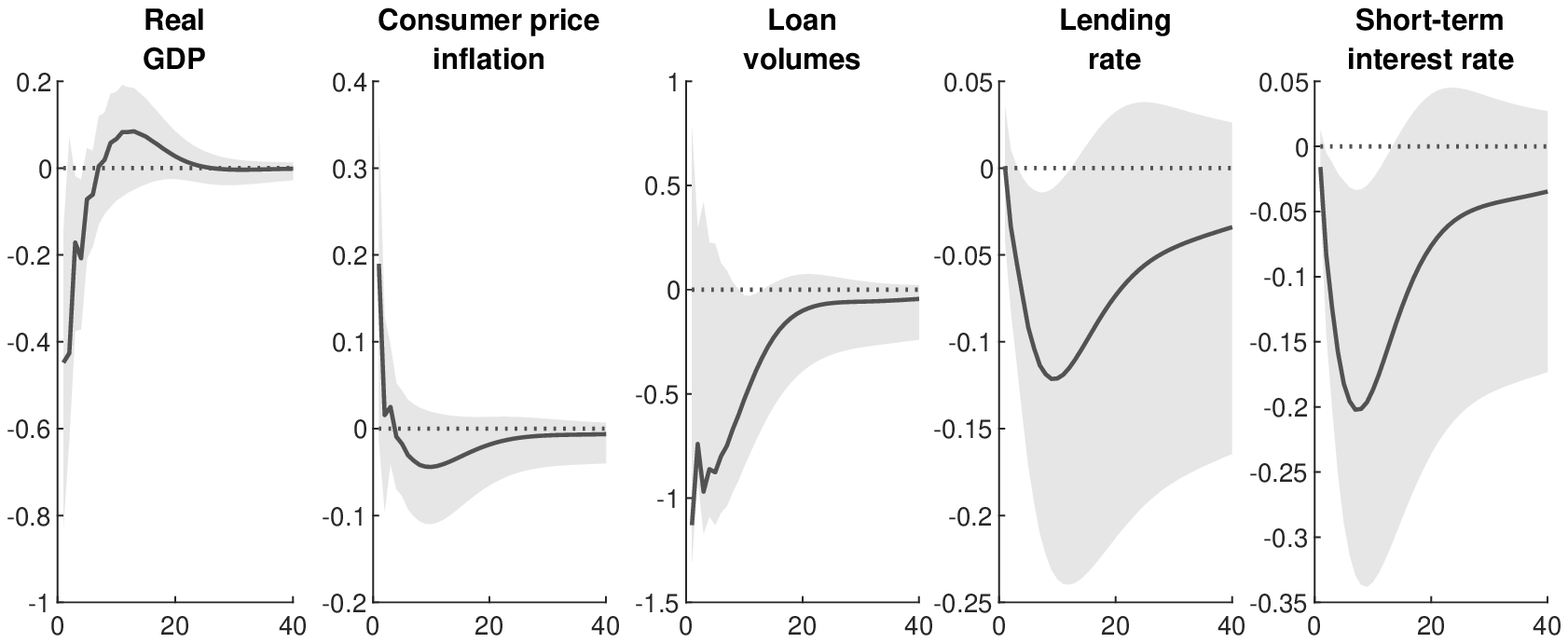}
\smallskip
\caption*{\textit{Notes}: See notes in \Cref{fig:irf2}.}
\label{fig:irf3}
\end{figure}

Next, we discuss the effects the macroeconomic effects of financial shocks by studying the impulse response functions to each shock. The impulse responses to demand and supply side financial shocks are presented in \Cref{fig:irf2} and \Cref{fig:irf3}, respectively. The solid line is the median and the shaded areas around the median denote the 68\% credible sets. \Cref{fig:irf2} shows that a contractionary demand side financial shock has a negative effect on real GDP and inflation. Real GDP growth falls on impact but the effect of the shock largely dissipates after three quarters. The effect on inflation is more pronounced. Moreover, inflation also decreases immediately, after which the effect gradually begins to dissipate. Loan volume growth tends to decrease less than growth of real GDP. Loan growth decreases immediately following the shock, but recovers within eight quarters following the shock. The short-term interest rate and the lending rate falls for five quarters following the shock, after which it begins to normalize slowly. 

As shown in \Cref{fig:irf3}, a supply side financial shock also leads to an immediate decline in the real GDP growth and inflation, after which the effects begins to dissipate. Real GDP growth turns positive 8 quarters following the shock, whereas the effect on inflation disappears after a year. Moreover, the decrease in loan volume growth is greater than the immediate effect on real GDP growth. Loan volumes decrease on impact and the effect gradually dissipates after 20 quarters. The lending rate and the short-term interest rate do not immediately respond to the shock but begin to gradually fall, reaching their lowest point after two years, and then slowly recovering from then onwards.

Despite the opposing sign of the effect on inflation, the response dynamics to supply side financial shocks are similar to those obtained in related studies, for example \cite{Gambetti2017} and \cite{Mumtaz2018}. However, in contrast to these studies, we observe both a larger and more persistent effect on loan volumes. On the other hand, demand side financial shocks have a larger effect on inflation in relation to changes in output and the effect on loan volume growth tends to be smaller compared to what \cite{Gambetti2017} finds.

\begin{figure}[H]
\caption{Forecast error variance decompositions for a supply side financial shock}
\smallskip
\includegraphics[scale=1]{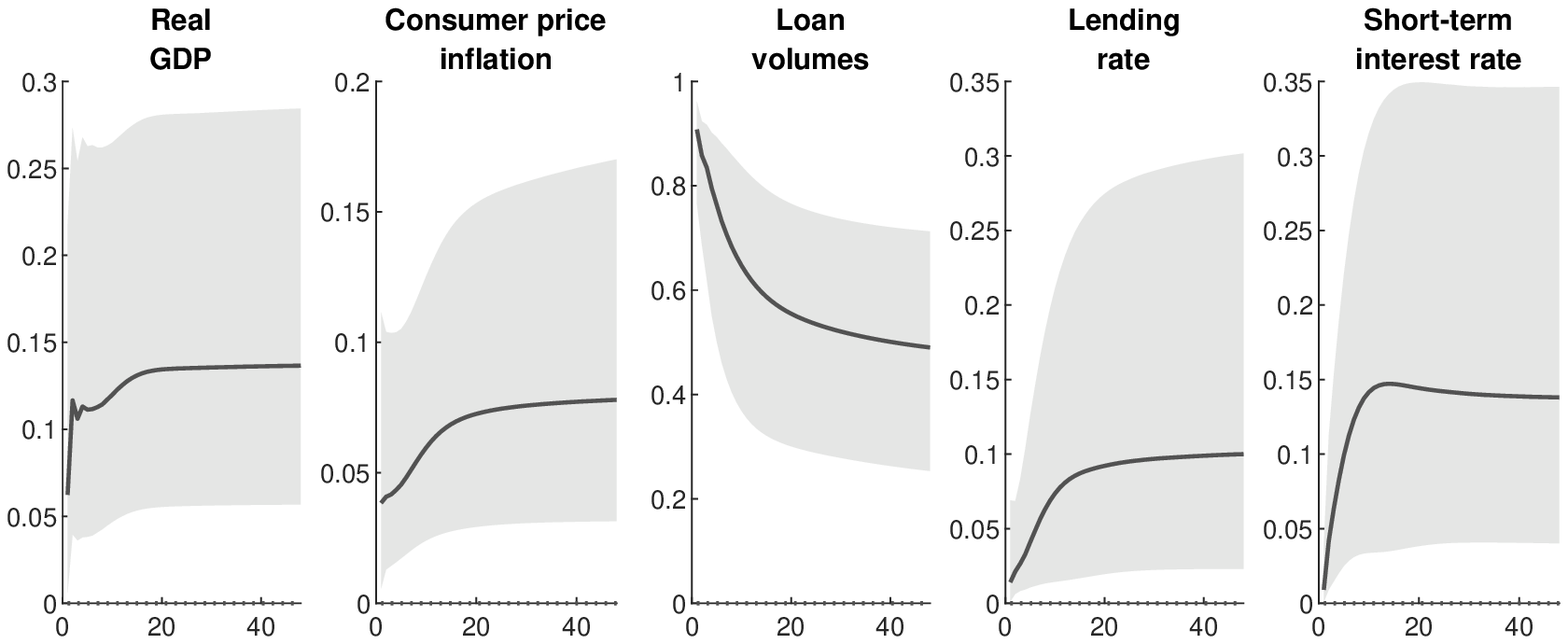}
\smallskip
\caption*{\textit{Notes}: The solid line is the posterior median and the shaded area is the 68th percentile posterior probability. Model is a VAR(2) including a constant. The estimation period is 1980:Q1-2011:Q4}
\label{fig:fevd3}
\end{figure}

The forecast error variance decomposition of the supply side financial shock is presented in \Cref{fig:fevd3}. It shows that supply side financial shocks initially account for roughly 90 percent of variance in loan volume growth and more than 50 percent at the four-year horizon. The contribution to forecast error variance of loan growth at four years' horizon is larger than \citepos{Hristov2012} and \citepos{Abbate2016} estimate of 10 and 44 percent, respectively. Moreover, we find that the contribution to the forecast error variance of GDP growth is roughly 10 percent. The finding is consistent with the estimates in other empirical studies that report a contribution to the forecast error variance of output between 5 and 20 percent \citep{Abbate2016, Gilchrist2012, Hristov2012, Meeks2012, Mumtaz2018}. The effect on inflation is also consistent with earlier studies. The contribution of supply side financial shocks to consumer price inflation is between 5 and 10 percent at the four-year horizon, which is in the range of 5 to 37 percent commonly reported in the literature \citep[see e.g.,][]{Abbate2016, Hristov2012, Gilchrist2012, Mumtaz2018}.

As discussed above, the inequality constraints for supply side financial shocks are also consistent with aggregate supply shocks as defined in \cite{Gambetti2017}, who only restrict aggregate supply shocks to have opposing effects on prices and output. Therefore, there is a legitimate concern that financial shocks may be misinterpreted to be aggregate supply shocks. To provide more evidence in support of the identification of the shocks, we turn to forecast error variance decompositions for additional information.

The supply side financial shock accounts for a large share of the variation in loan volume growth at all horizons (see \Cref{fig:fevd3}). This observation may be interpreted to support of the notion that innovations in loan volume growth originate in the financial sector and  may therefore also ease worries that the identified supply side shocks are mistaken for aggregate supply shocks\footnote{In a comparison of identification schemes using a Monte Carlo experiment, \cite{Mumtaz2018} find that sign restrictions combined with a condition that the identified shock maximizes forecast error variance of the quantity of loan supply up to 40 quarters, along with a proxy SVAR identification scheme, best match the impulse responses obtained from a DSGE model with financial frictions}.

\begin{figure}[H]
\caption{Forecast error variance decompositions for a demand side financial shock}
\smallskip
\includegraphics[scale=1]{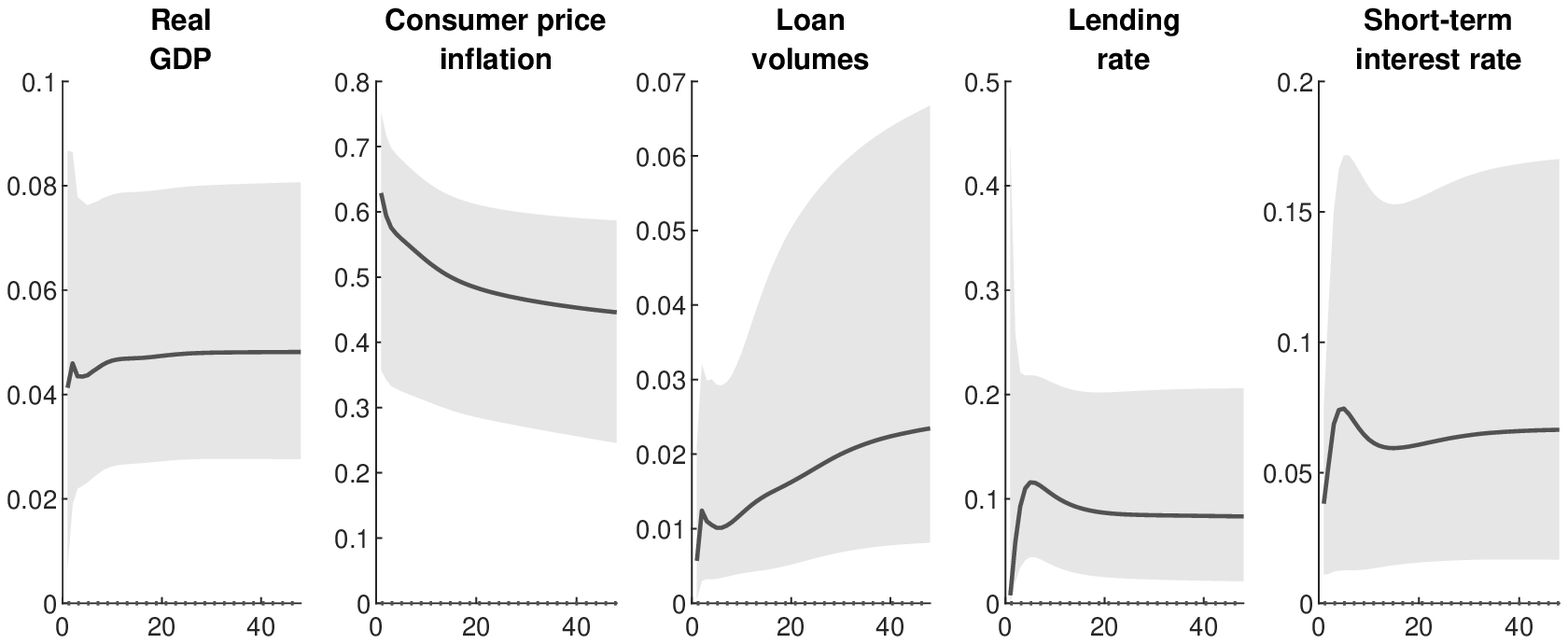}
\smallskip
\caption*{\textit{Notes}: See notes in \Cref{fig:fevd3}.}
\label{fig:fevd2}
\end{figure}

\Cref{fig:fevd2} presents the forecast error variance decomposition for the demand side financial shock. The demand side financial shock accounts for a large share of the variation in inflation but only a small share of the variance of other variables. The contribution of demand side financial shocks to the forecast error variance of consumer prices is roughly 50 percent at the four-year horizon. The effect on inflation is considerably larger than reported in the previous empirical literature on financial shocks. However, the findings are consistent with the empirical literature on the macroeconomic effect of financial shocks. The relatively small contribution to real GDP is consistent with \citepos{Blanchard1988} interpretation that demand side shocks have only temporary effect on output. Moreover, \cite{Forni2010} find that a demand side shock explains approximately 50 percent of the forecast error variance of the GDP deflator at the six-year horizon and under 8 percent of the forecast error variance of the GDP. 

\begin{figure}[H]
\caption{Historical decomposition of loan volume growth}
\smallskip
\includegraphics[scale=1]{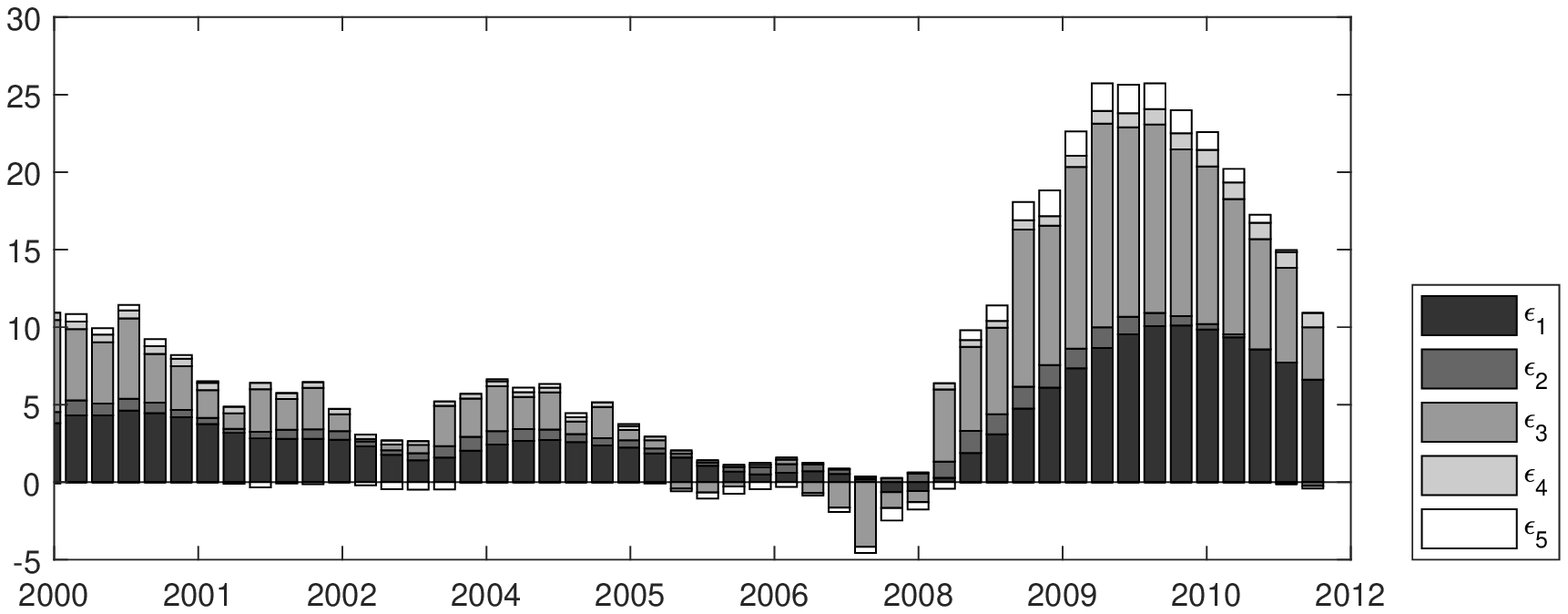}
\label{fig:hd3}
\end{figure}

\begin{figure}[H]
\caption{Historical decomposition of consumer price inflation}
\smallskip
\includegraphics[scale=1]{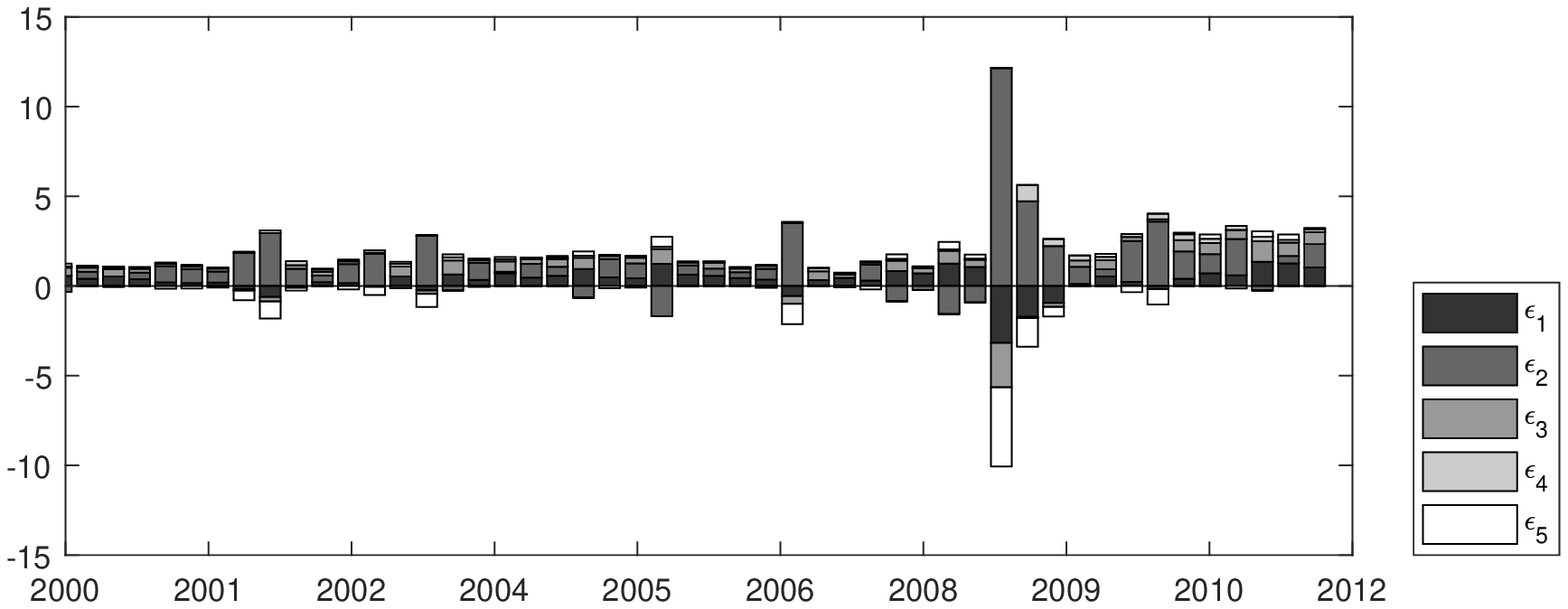}
\label{fig:hd2}
\end{figure}

Next, we discuss the contribution of the shocks to the evolution of consumer price inflation and loan volume growth using historical decompositions. The historical decompositions for loan volume growth and consumer price inflation are shown in \Cref{fig:hd3} and \Cref{fig:hd2}, respectively. %

\begin{figure}[H]
\caption{Historical decomposition of GDP growth}
\smallskip
\includegraphics[scale=1]{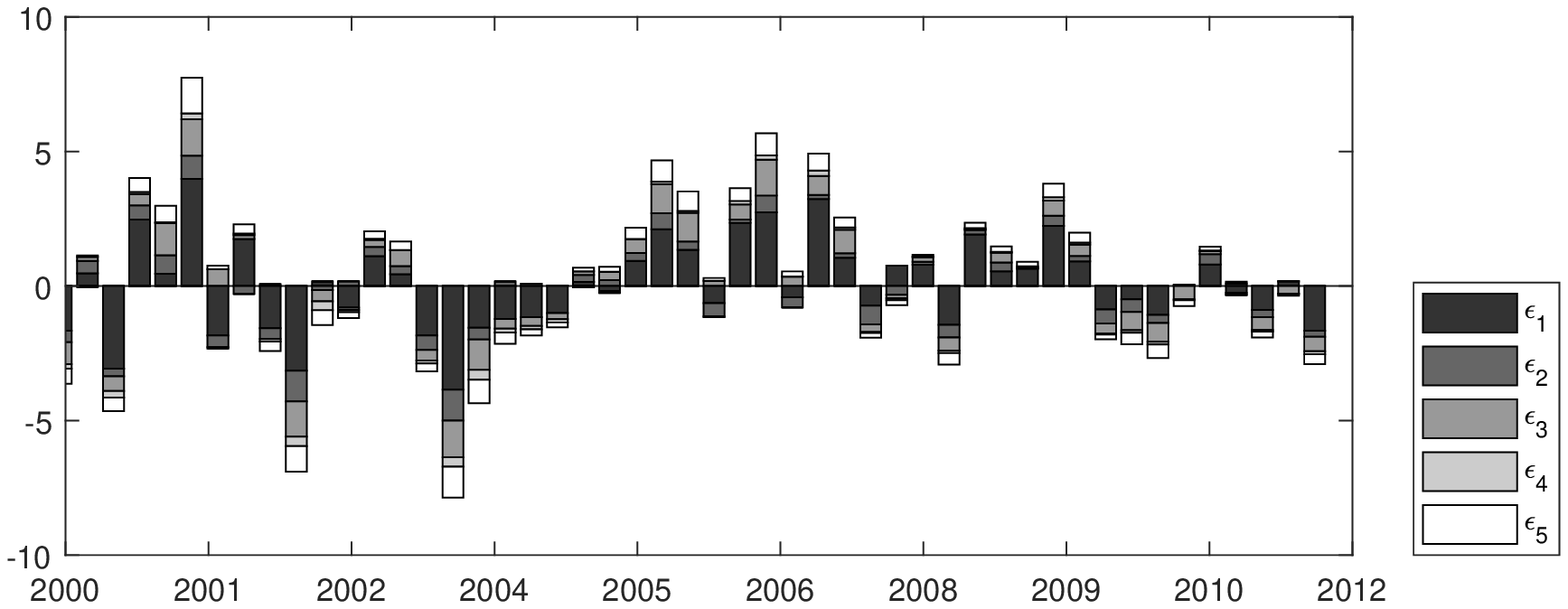}
\label{fig:hd1}
\end{figure}

As can be seen in \Cref{fig:hd3}, the supply side financial shock ($\epsilon_3$) explains the largest share of the fall in loan volume growth at the onset of the financial crisis and contributes most to the increase in loan growth during the recovery. Furthermore, the demand side financial shock seems to have contributed only marginally to the increase in loan volume growth during the beginning of the recovery period. 

\Cref{fig:hd3} shows that both financial shocks have contributed positively to inflation during the recovery period, although the cumulative contribution of supply side financial shocks to inflation is small in relation to demand side financial shocks. Nevertheless, the positive contribution to inflation supports the argument that financial shocks have attenuated the disinflationary pressures in the aftermath of the financial crisis \citep{Gilchrist2017}.

\section{Conclusion} \label{sec:conclusion}

In this paper, we study the effects of financial shocks on inflation in the United States. We exploit the non-Gaussian features in the time series to uniquely identify the SVAR model, which allows us to assess the inflation effects of financial shocks. 

Our results show that the data is supportive of two distinct financial shocks that have opposite inflationary responses. This suggest that a financial shock is transmitted to the real economy via both aggregate demand and aggregate supply channels. This finding is important from a policy perspective, given that a cost-push shock, which moves output and inflation in opposite directions, poses a trade-off with the central banks' objective of price and output stability.  This finding also has implications for the theoretical work on the transmission of financial shocks. Theoretical models that include the financial sector would benefit from the inclusion of both demand and supply side effects of financial shocks to better capture the effects on the real economy and the impact of monetary policy. Nevertheless, historical decompositions suggest that the effect of both financial shocks on consumer prices is small. Therefore, negative financial shocks alone do not account for the lack of disinflation in the US during the Great Financial Crisis. Moreover, given that financial shocks seem to exhibit both inflationary and disinflationary effects, as also argued by \cite{Gerali2010}, empirically tracing the origins of these different types of financial shocks poses an interesting avenue for further research.

\clearpage
\section*{Acknowledgements} \label{sec:acknowledgements}

I gratefully acknowledge financial support from Academy of Finland (grant 308628), the Yrj{\"o} Jahnsson Foundation (grant no. 6609, 7016), the Foundation for the Advancement of Finnish Securities Markets, and the Nordea Bank Foundation. I would also like to thank Markku Lanne and Jani Luoto for their advice in the preparation of the manuscript.

\bibliographystyle{apalike}
\bibliography{bib3}  

%
%
%
%

\end{document}